\documentclass[prc,aps]{revtex4}
\usepackage{graphicx}
\begin{document}

\title{Collective behavior in random interaction
\footnote{Contribution to XXIX Symposium on Nuclear Physics, Cocoyoc, Mexico, 2006}}

\author{Calvin W. Johnson and Hai Ah Nam}
\affiliation{Physics Department, San Diego State University, 5500
Campanile Drive, San Diego, California 92182-1233}

\begin{abstract}
Recent investigations have looked at the many-body spectra of random two-body
interactions. In fermion systems, such as the interacting shell
model, one finds pairing-like spectra, while in boson systems, such
as IBM-1, one finds rotational and vibrational spectra. 
We discuss the search for random ensembles of fermion
interactions that yield rotational and vibrational spectra, and in
particular present results from a new ensemble, the ``random
quadrupole-quadrupole ensemble''
\end{abstract}
\maketitle

\section{Introduction: the puzzle of collectivity}

Even-even nuclei have long been known to exhibit a wide range of
collective behavior. The broadest
classification of collective behavior is into three groups:
pairing, vibrational, and rotational collectivity\cite{BM,simple}. The first is in
analogy with superconductivity in metals, and the latter two based
upon quadrupole deformations of a liquid drop.

The spectral signatures of collectivity are:

\noindent (1) The quantum numbers of the ground state. In particular
for pairing one expects $J^\pi_\mathrm{g.s.}=0^+$.

\noindent (2) Regularities in the excitation energies, in particular
among the lowest $J^\pi = 2^+, 4^+, 6^+, \ldots$ states relative to
the $J^\pi = 0^+$ ground state. Particularly useful measures are the ratios
$R_{42} \equiv E_x(4_1^+)/E_x(2^+_1)$ and $ R_{62} \equiv
E_x(6_1^+)/E_x(2^+_1)$. For pairing, vibrational, and rotational
collectivity one expects $R_{42} = 1,2,$ and $3.33$, respectively.

\noindent (3) Strong, correlated B(E2) transition strengths, which
measure the collectivity of the wavefunctions. (In particular one
expects strong intraband transitions and weak interband transitions,
but we will not consider that further here.) Traditionally one
either compares B(E2) strengths to the single-particle limit (Weisskopf 
units) or ratios of B(E2)s.

Simple and successful models of collectivity, usually based upon
group theory, are well known \cite{simple} ; in fact more sophisticated
classifications  are possible and are used, but we do not consider them here. 

In order to understand the roots of collectivity, start from the coordinate-space Hamiltonian
\begin{equation}
\hat{H} = \sum_i -\frac{\hbar^2}{2M_N} \nabla_i^2 + \sum_{i < j}
V(\vec{r}_i, \vec{r}_j).
\end{equation}
with a chosen nucleon-nucleon potential. Although there are many
ways to solve this Hamiltonian, we consider the shell-model route:
one chooses a single-particle basis $\left \{\phi_i(\vec{r})\right
\}$ and writes the Hamiltonian in occupation space using second
quantization, where $\hat{a}^\dagger_i$ creates a particle in the 
single-particle state $\phi_i$,
\begin{equation}
\hat{H} = \sum_i \epsilon_i \hat{a}_i^\dagger \hat{a} + \frac{1}{4}
\sum_{ijkl} V_{ijkl} \hat{a}_i^\dagger \hat{a}^\dagger_j \hat{a}_l
\hat{a}_k .
\end{equation}
The $\epsilon_i$'s are the single-particle energies and the
$V_{ijkl}$'s are the two-body matrix elements, in principle 
computed from antisymmeterized integrals of the nucleon-nucleon
potential:
\begin{equation}
V_{ijkl} = \int d^3r d^3r^\prime \phi_i^*(\vec{r}) \phi_j^*
(\vec{r}^\prime) V(\vec{r},\vec{r}^\prime) \left [ \phi_k(\vec{r})
\phi_l(\vec{r}^\prime) -\phi_l(\vec{r}) \phi_k(\vec{r}^\prime) \right
]
\end{equation}

Any interacting shell-model code, such as the
one we use, REDSTICK\cite{redstick}, reads in: the list
of valence single-particle states (such as the
$1s_{1/2}$-$0d_{3/2}$-$0d_{5/2}$ or $sd$ space or the
$1p_{1/2}$-$1p_{3/2}$-$0f_{5/2}$-$0f_{7/2}$ or $pf$ space);
information on the \textit{many-body} space, such as 
valence protons and neutrons; and finally a list
of the single-particle energies and two-body matrix elements, both
of which are read in as numbers. The two-body matrix elements also conserve angular
momentum $J$ and frequently, but not always, isospin $T$.

As mentioned, there exist simple algebraic models of
collectivity. Importantly, they have representations both in fermion
and boson spaces, with the two-body matrix elements derived from the
group generators. Furthermore, these algebra-based models
\textit{describe well much of the experimental spectra
of even-even nuclides}. Therefore, the default assumption has long
been that the data in turn imply that the real nuclear interaction
must have buried deep inside strongly ``algebraic'' components. 
This assumption can be tested by sampling a large number of 
Hamiltonians and seeing if any signatures of collectivity arise \cite{JBD98}.

\section{The two-body random ensemble (TBRE)}

The simplest test is to replace the two-body matrix elements with
random numbers, independent except that angular momentum $J$ (and
isospin $T$) is (are) conserved. This is known as the two-body random
ensemble (TBRE) and, as long as the ensemble is symmetric about
zero, the results are broadly insensitive to the details of the
distribution\cite{Jo99}, e.g. weighting with $J$, uniform or Gaussian
distribution, etc.  (The displaced TBRE, or DTBRE\cite{dtbre}, which is
\textit{not} symmetric about zero, will be discussed briefly below.)
The TBRE was originally used to investigate quantum chaos\cite{WF72}, and so
such investigations considered only states with the same quantum
numbers. It was not until a few years ago that investigations 
compared spectral properties of states with different quantum
numbers\cite{JBD98}.

Use of the TBRE is straightforward. A sample interaction from the
ensemble is generated, and fed into a shell-model diagonalization
code. For the TBRE this means that $V_{JT}(ij,kl) = $ an independent random
number. The many-body system is referenced by its nuclear physics 
analog, e.g., ``$^{22}$O'' mean six ``neutrons'' (identical fermions) 
in an $sd$ valence space, although the actual many-body space is 
abstracted and is no longer tied to, for example, 
the radial wavefunctions.  The output (for example, the angular momentum of the ground
state) is stored in a file for later analysis. Typically one compiles 
several hundred or even several thousand runs to generate
good statistics.

Previous investigations into the properties of the TBRE in fermion
systems yielded the following results:

\noindent $\bullet$ Dominance of ground states with $J^\pi=0^+$. (Hereafter we 
will drop parity, as most of the model spaces investigated do not have 
abnormal parity states.)
Specifically, in many-body spaces where the fraction of $J=0$ states is
small, typically 4-11 \%, after diagonalization of Hamiltonians
drawn from the TBRE, the fraction of ground states with $J=0$ is
45-70\%. This is the best known result from random interactions.

\noindent $\bullet$ Pairing-like behavior\cite{JBDT00}. In addition to $J=0$
ground states, one also finds odd-even staggering of ground state
binding energies, and a ground state ``gap.''

\begin{figure}
%\begin{center}
\includegraphics[scale=0.5]{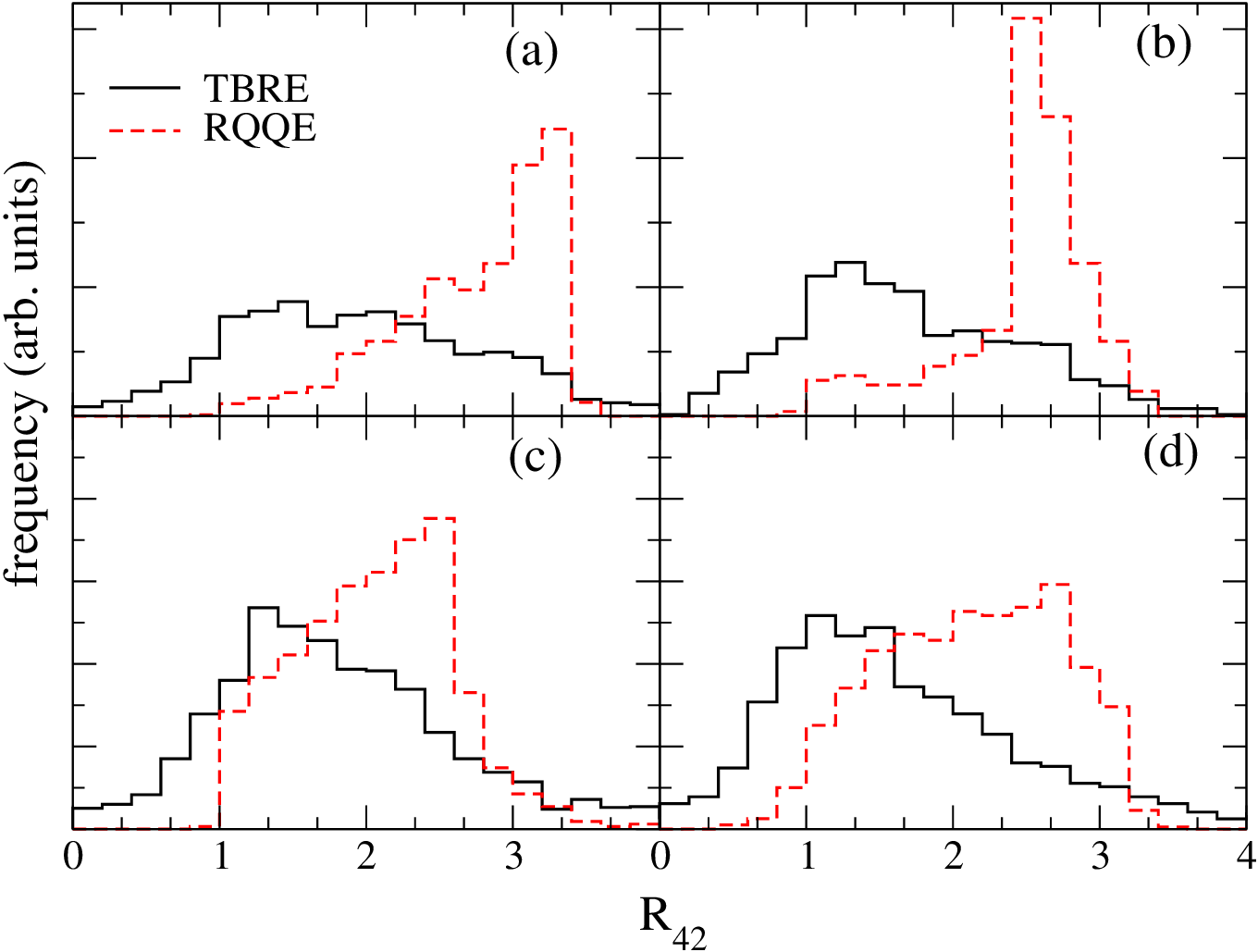}
%\end{center}
\caption{Probability distribution of $R_{42}$ for TBRE (solid line) and RQQE 
(dashed line) for (a) $^{24}$Mg; (b) $^{28}$Si; (c) $^{22}$O; (d) $^{48}$Ca.}
\end{figure}

For a pairing system, one expects $R_{42} \sim 1$. 
Fig.~1 shows the distribution of $R_{42}$ over a thousand samples from the TBRE 
for several different many-fermion systems. All are peaked near 1 (pairing), but 
with broad distributions that include $R_{42} = 2$ (vibration) and 3.33 (rotational). 

Many-boson systems, in particular the IBM-1, have also been studied using 
the TBRE, also yielding a dominance of $J=0$ ground states\cite{BF00}.
In contrast to fermion systems, however, the distribution of $R_{42}$ is
sharply peaked at 2 and 3.33.  Bijker and Frank 
studied the ratio
$\mathrm{B(E2:4_1^+ \rightarrow 2_1^+)/B(E2:2_1^+ \rightarrow
0_1^+)}$, which turns out for IBM-1 to be strongly correlated with $R_{42}$ 
and agrees with analytic algebraic limits; plots of such correlations 
we will call Bijker-Frank plots. This is interpreted as
evidence for band structure in bosons systems with random
interactions. 
The boson results have been successfully explained using a
mean-field analysis\cite{BF01}.  

There have been numerous papers ``explaining''
the fermion results; for reviews see\cite{ZAY04,ZV04}.
Much of the focus has been on the predominance
of $J=0$ ground states; few have considered 
band structure beyond $J=4$, save for a few exceptions \cite{JBD98,ZV04}.  
In Fig.~2 we give a correlation plot for for $R_{62}$ vs.
$R_{42}$. Fig.~2(a) shows results  using ``realistic'' interactions
calculations in the $sd$ \cite{Wi84} and
$pf$ \cite{gxpf1} shell; 2(b)-(d) are results for the TBRE in several different 
fermion systems.
Somewhat surprisingly, we get a strong correlation that follows
exactly the naive predictions for pairing, vibrational, and
rotational collectivity. (Although we do not show it, this 
result does not appear in single-$j$ systems frequently used as 
testbeds for random interactions). This higher-order band structure is a challenge to any
claim to understanding random
interactions results for fermion systems. Although we do not show it,
similar (and sharper) results occur in the IBM with random
interactions.

\begin{figure}
\begin{center}
\includegraphics[scale=0.5]{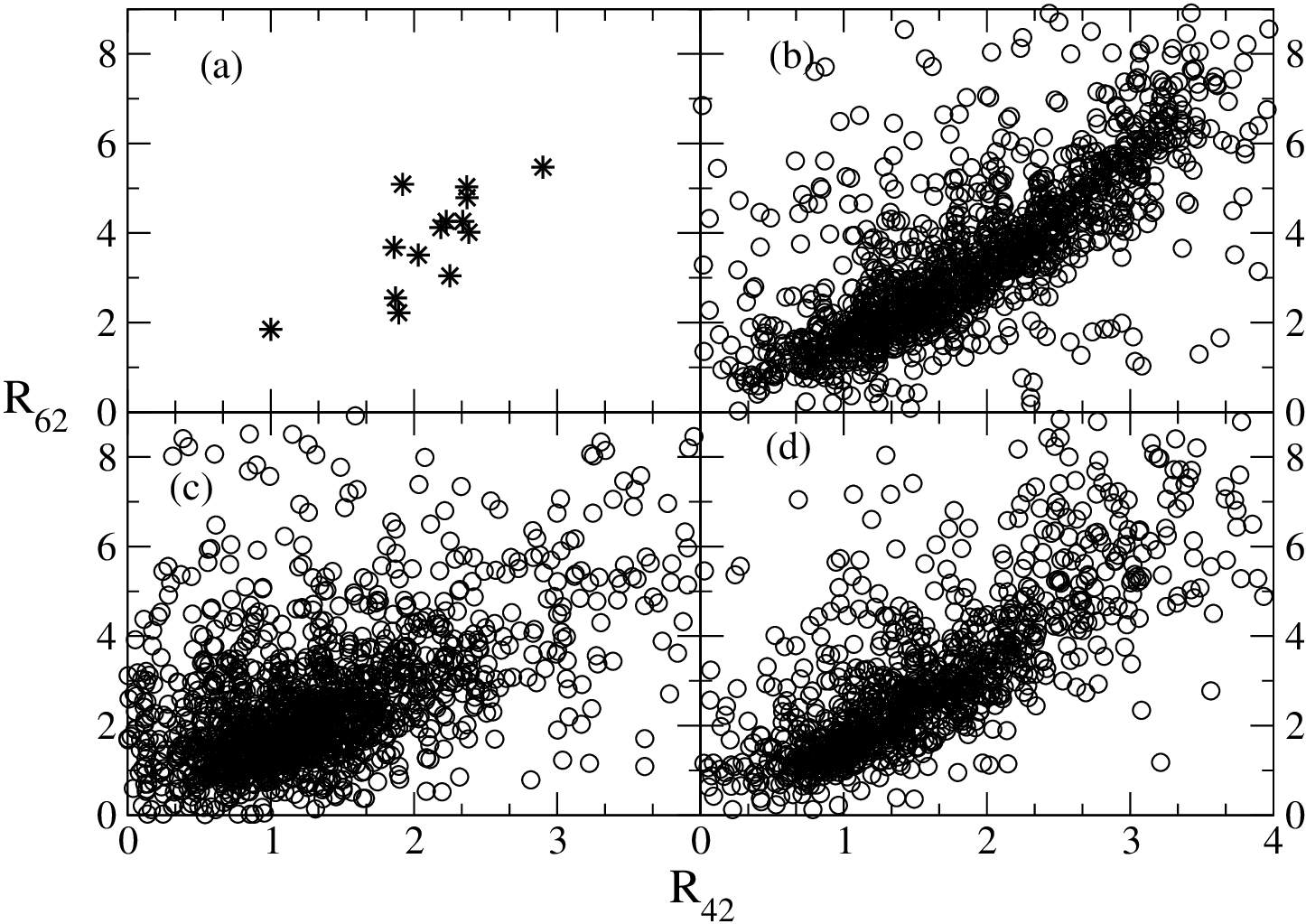}
\end{center}
\caption{Correlation of $R_{62}$ vs. $R_{42}$ (a)
"realistic" calculations in $sd$ and
$pf$ shells; TBRE in (b) $^{24}$Mg; (c) $^{44}$Ti; (d) $^{48}$Ca.}
\end{figure}

\begin{figure}
\begin{center}
\includegraphics[scale=0.5]{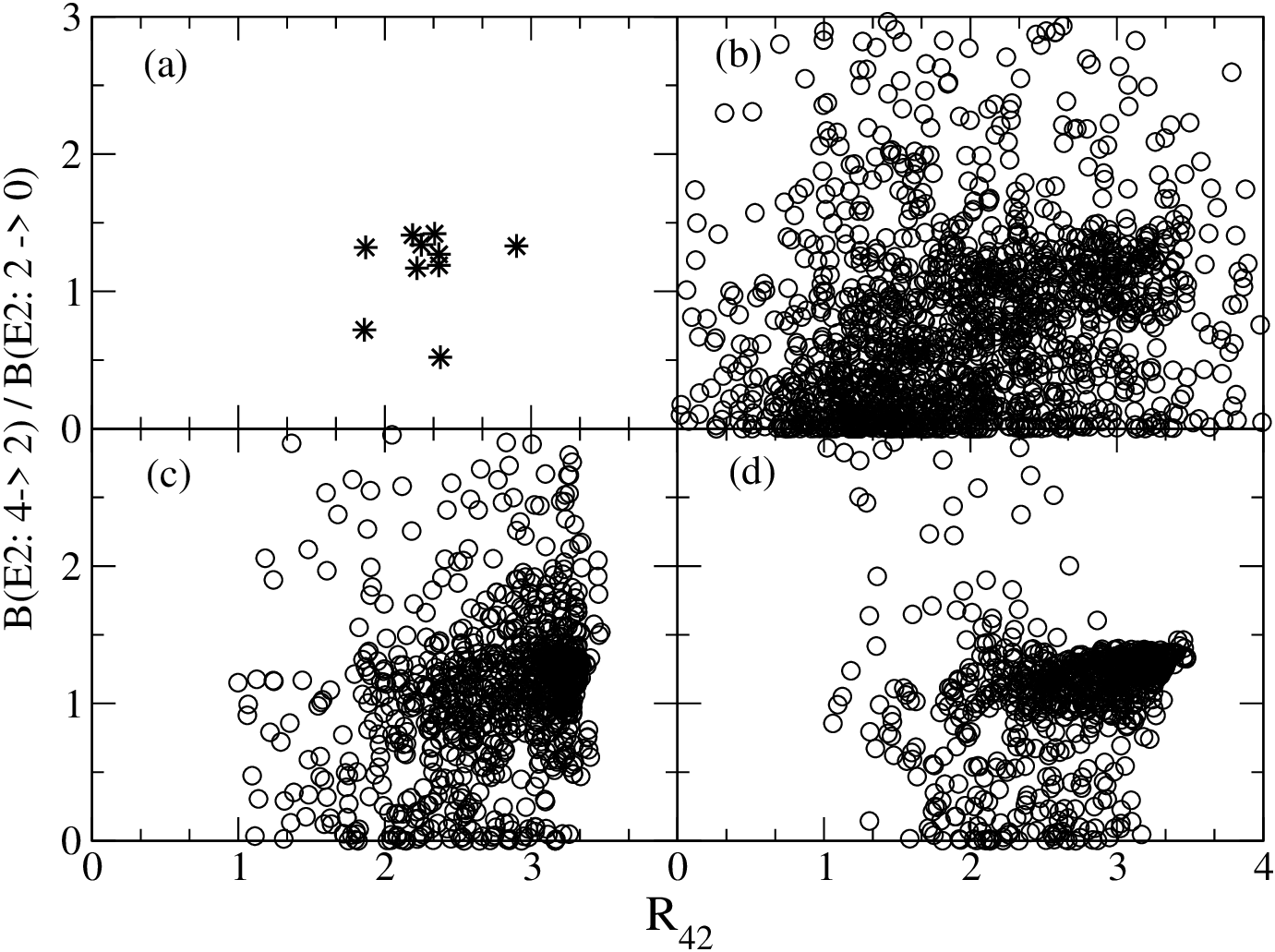}
\end{center}
\caption{Bijker-Frank plots for fermion systems. (a) ``Realistic''
systems in the $sd$ and $pf$ shell. (b)-(d) for $^{24}$Mg: (b) TBRE. (c) RQQE with standard
quadrupole transition operator. (d) RQQE with `consistent'
transition operator.}
\end{figure}

Fig.~3 shows correlations of the ratio B(E2:$4\rightarrow 2$)/B(E2:$2\rightarrow 0$) 
vs. $R_{42}$ (Bijker-Frank plots), for ``realistic'' $sd$ and 
$pf$ shell calculations in Fig.~3(a), and for the TBRE in $^{24}$Mg in
Fig.~3(b). (For the B(E2) transitions in TBRE, for lack of other 
constraints we assume harmonic-oscillator radial wavefunctions.) 
While IBM-1 Bijker-Frank plots show sharp correlations, 
the fermion TBRE does not show very strong correlations. 
(Another approach to collectivity in fermion systems with random 
interactions is the Alaga ratio\cite{ZV04}).

Fig.~4 shows correlations between B(E2:$6\rightarrow 4$)/B(E2:$2\rightarrow 0$) 
and B(E2:$4\rightarrow 2$)/B(E2:$2\rightarrow 0$), which has not 
been previously studied. Again, Fig.~4(a) shows realistic results, 
while Fig.~4(b) is for the TBRE. The correlations are sharper than 
for the Bijker-Frank plot. Not shown are results for IBM-1, which  
has still sharper correlations.  

To summarize: fermion systems with TBRE interactions show some 
signatures of collectivity, but much more weakly than boson systems. 
Why is not understood. This motivates the next section.

\section{The random quadrupole-quadrupole ensemble (RQQE)}

One way to probe the difference between random interactions in boson
and fermion systems is to search for a random ensemble that yield
stronger collective structure in fermion systems. One proposal is
the displaced TBRE, or DTBRE, where the two-body matrix elements are
given a constant displacement. While an appealing suggestion, 
we have found that the
DTBRE displays collective behavior only for a handful of even-even systems. Even worse,
as discussed below, the resulting ensemble is not very
``random.''

Therefore, inspired by the ``consistent-Q'' formulation in the
interacting boson model\cite{BF00}, we propose the following. Consider the
general one-body operator of angular momentum rank 2:
\begin{equation}
\tilde{Q}_m = \sum_{ij} q_{ij} \left [ \hat{a}_i^\dagger \times
\hat{a}_j \right ]_{2m}.
\end{equation}
If $q_{ij} = \langle i || r^2 Y_2 || j \rangle$, then $\tilde{Q}$ is
the standard quadrupole operator. But instead we choose the $q_{ij}$
randomly. We then define the \textit{random quadrupole-quadrupole
ensemble} to be interactions
\begin{equation}
\hat{H} = -\lambda \tilde{Q} \cdot \tilde{Q}.
\end{equation}
The antisymmeterized matrix elements are $V_{ijlk} = q_{ik}q_{jl} -
q_{il} q_{jk}$ (here and above we have left out details of the
angular momentum coupling which are straightforward, if tedious, to
include). There is also a one-body single-particle energy induced by
normal ordering of the operators.

Any member interaction of the RQQE has fewer random parameters than
the TBRE. For example, in the $sd$ shell, the TBRE has 63
independent, randomly generated two-body matrix elements, while the
RQQE, if one assumes time-reversal symmetry (that is, 
$q_{ab} = (-1)^{j_a - j_b} q_{ba}$), has only 5 independent
parameters; for the $pf$ shell, the TBRE has 195 independent matrix 
elements and the RQQE has 9. 

The RQQE has $J=0$ ground states $> 99\%$ for \textit{all} even-even nuclides 
we tried. In Figs.~1-4 we show other results for the RQQE alongside those for the TBRE. 
To summarize them:

\noindent $\bullet$ In Fig.~ 1 we see $R_{42}$ for the RQQE peaks between 2 and 3.3, that is, somewhere 
between vibrational and rotational. 

\noindent $\bullet$  Although not shown, the correlation of $R_{62}$ vs. $R_{42}$ for 
the RQQE is similar to that of the TBRE  in Fig.~2.

\noindent $\bullet$  In Fig.~3(c)-(d), the Bijker-Frank plot shows the correlation between ratios of 
B(E2)s and $R_{42}$. Unlike in the TBRE, the RQQE shows much stronger correlation. Fig.~3(c) 
uses the ``standard'' quadrupole transition operator, assuming harmonic oscillator single-particle 
states; Fig.~3(d) uses a ``consistent'' quadrupole transition operator, the same $\tilde Q$ 
used in the Hamiltonian. Both results are similar, although the consistent-quadrupole results are 
sharper.

\noindent $\bullet$ In Fig.~4(c)-(d) we show correlations in B(E2)s in the yrast band 6-4-2-0. 
The correlations are much sharper than for the TBRE, and sharpest for the consistent-quadrupole 
calculation.

\begin{figure}
\begin{center}
\includegraphics[scale=0.5]{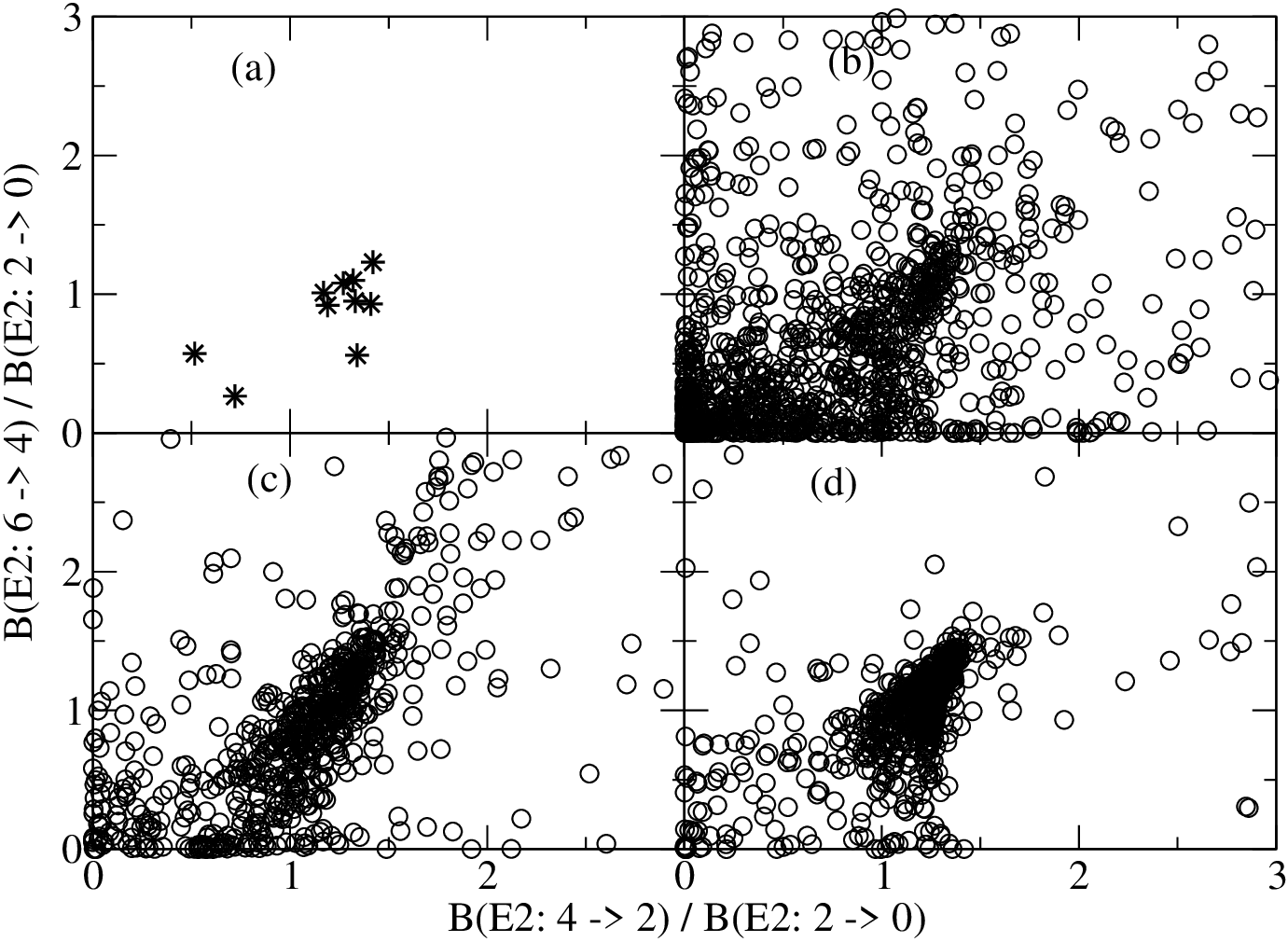}
\end{center}
\caption{Correlation of B(E2) ratios B(E2: 6 $\rightarrow$ 4) / B (E2: 2 $\rightarrow$ 0)
versus B(E2: 6 $\rightarrow$ 4)/B(E2: 2 $\rightarrow$ 0) for
(a) ``Realistic''
systems in the $sd$ and $pf$ shell. (b)-(d) for $^{24}$Mg: (b) TBRE. (c) RQQE with standard
quadrupole transition operator. (d) RQQE with `consistent'
transition operator.}
\end{figure}

\section{Test for randomness}

Although the RQQE shows stronger spectral signatures for band structure in fermion systems than 
the TBRE, because the former has many fewer random parameters than the latter, we are 
concerned if the RQQE results are truly `random.'  We propose the following as a necessary, 
if not sufficient, test of 
randomness.

For any given many-fermion system, let $N_0$ be the dimension of the $J=0$ subspace. 
Therefore, any $J=0$ wavefunction can be represented as a vector of unit length in 
an $N_0$-dimensional space. If we choose the vectors randomly, then they will uniformly 
cover a unit sphere, and it is straightforward to show that the angle $\theta$ between 
any two vectors will have a probability $\propto \sin^{N_0-2}\theta$. 

We compared ground state vectors for $J=0$ ground state. (We ignored interactions 
that did not have have $J=0$ ground states. ) We took the dot product between two randomly chosen ground state 
vectors, computed the angle between them, and binned the results, 
shown in Fig.~5 for $^{48}$Ca. The expected $\sin^{(N_0 -2)} \theta$ is shown for comparison.

Fig.~5(a) shows the distribution for the fermion TBRE. It follows exactly the expected 
$\sin^{(N_0 -2)} \theta$ form. 

Fig.~5(b) shows the distribution for RQQE. It is closer to the expected $\sin^{(N_0 -2)} \theta$ 
than the DTBRE in (c) and (d) but is not completely random. The distortion may be due to the low number of 
independent random parameters.

Fig.~5 (c) and (d) shows the distribution for the displaced TBRE (DTBRE) for displacements $c \approx 1, 3$, 
respectively.  
(Following the original paper, the width of the displaced Gaussian is 0.6.) 
Note as displacement $c$ gets larger, the distribution moves further from $\sin^{N_0 -2} \theta$
and becomes more peaked towards $\theta = 0$. These ground state wavefunctions are \textit{not} 
randomly distributed; rather they are clustered about the limit wavefunction for $c \rightarrow \infty$.

This is strong evidence that TBRE wavefunctions are 
randomly and uniformly distributed; it is also evidence that the DTBRE wavefunctions 
are \textit{not} uniformly distributed. The results from RQQE are inconclusive; they are not uniformly 
distributed as the TBRE wavefunctions are, but they are ``more random'' than DTBRE.

\begin{figure}
\begin{center}
\includegraphics[scale=0.5]{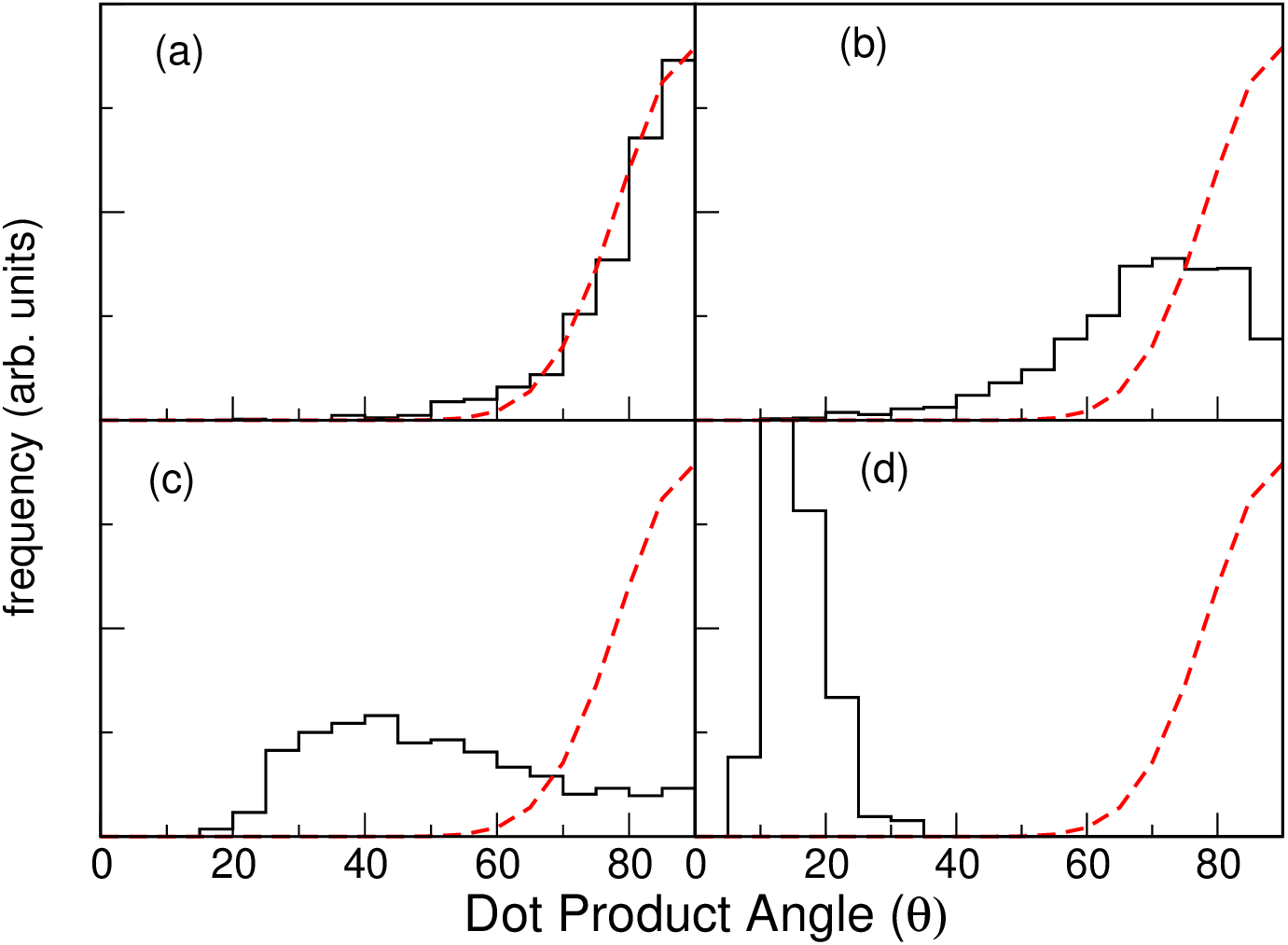}
\end{center}
\caption{Distribution of angles between different 
$J=0$ ground state wavefunctions 
(for $^{48}$Ca) for (a) TBRE; (b) RQQE; (c) DTBRE with displacement 
$c = 1$; (d) DTBRE with displacement $c$ = 3. Dashed line is 
$\sin^{(N_0 -2)} \theta$, where $N_0$ is the dimension of the $J = 0$ 
subspace. }
\end{figure}

\section{Conclusions}

We have revisited the question of signatures of collectivity in the spectra of 
random, two-body interactions in fermion systems. Some surprising, previously 
unknown correlations show up even in the two-body random ensemble (TBRE). 
We proposed a new ``random'' ensemble, 
the random quadrupole-quadrupole ensemble (RQQE), and found it had significantly sharper 
signatures of collectivity than the TBRE. While this does not solve the puzzle of collectivity,
it does demonstrate that typical spectral signatures of collectivity do not 
rigorously require the standard Elliot quadrupole-quadrupole or SU(3) interaction. 

 Of course, collective behavior is 
not surprising if the ``ensemble'' is not very random, and we looked at the dot 
product between ground state wavefunctions: the RQQE  deviates from the 
expected random distribution but much less so than the displaced TBRE or DTBRE.

The U.S.~Department of Energy supported this investigation through
grant DE-FG02-96ER40985.

\end{document}